# Signature of Dispersing 1D Majorana Channels in an Iron-based Superconductor

Zhenyu Wang[1], Jorge Olivares Rodriguez[1], Lin Jiao[1], Sean Howard[1], Martin Graham[2], G. D. Gu[3], Taylor Hughes[4], Dirk K. Morr[2], and Vidya Madhavan[1]

[1]Department of Physics and Frederick Seitz Materials Research Laboratory, University of Illinois Urbana-Champaign, Urbana, Illinois 61801, USA
[2] Department of Physics, University of Illinois at Chicago, 845 W. Taylor St, m/c 273, USA
[3] Condensed Matter Physics and Materials Science Department, Brookhaven National Laboratory, Upton, NY 11973, USA
[4] Department of Physics and Institute for condensed Matter Theory, University of Illinois at Urbana-Champaign, Urbana, Illinois 61801, USA

**The possible realization of Majorana fermions as quasiparticle excitations in condensed matter physics has created much excitement. Most recent studies have focused on Majorana bound states which can serve as topological qubits. More generally, akin to elementary particles, Majorana fermions can propagate and display linear dispersion. These excitations have not yet been directly observed, and can also be used for quantum information processing. One route to realizing this is in a line junction between two phase-shifted superconductors coupled to topological surface states. Recent theory indicates that in iron-based superconductors, a particular type of crystalline defect, i.e., a domain wall (DW) between two regions with a half-unit cell shift between them, should create a π-phase shift in the superconducting order parameter. Combined with recent data showing topological surface states in FeSe$_x$Te$_{1-x}$ we find that this is the ideal system to realize helical 1D-dispersing Majorana modes. Here we report scanning tunneling spectroscopic (STS) measurements of crystalline DWs in FeSe$_{0.45}$Te$_{0.55}$. By analyzing large-area superconducting gap maps, we identify the gap in the topological surface state, demonstrating that our sample is an effective Fu-Kane proximitized topological system. We further locate DWs across which the atoms shift by half a unit cell. STS data on these DWs reveal a flat density of states inside the superconducting gap, a hallmark of linearly dispersing modes in 1D. This unique signature is absent in DWs in the related superconductor, FeSe which is not in the topological phase. Our combined are consistent with the observation of dispersing Majorana states at a π-phase shift DW in a proximitized topological material.**

Majorana fermions are putative elementary particles that are their own anti-particles[1]. Emergent analogs of these fermions have been argued to exist as quasiparticle excitations in condensed matter systems[2-7], and have attracted much attention as possible building blocks of fault-tolerant quantum computation[8, 9]. So far, various predictions and realizations of localized Majorana bound states (MBS) have been reported. The platforms include strong spin-orbit coupled semiconductor nanowires[10-14], ferromagnetic atomic chains[15-17], and topological insulators that are proximity-coupled with s-wave superconductors[18, 19]; in all cases the MBS were spectroscopically determined as zero-energy conductance anomalies. In addition to the localized MBS, however, theoretical predictions show that dispersing Majorana states may also be realized as quasiparticles in condensed matter systems[18, 20, 21]. These quasiparticles are of fundamental interest, and may be harnessed for quantum computing. For example, 1D chiral Majorana modes have been proposed in hybrid systems combining superconductors with a quantum anomalous Hall insulator[22] or 2D magnetic Fe islands[23, 24]. However, most of these platforms are complex heterostructures that are difficult to fabricate, and are stable only at low temperatures. This makes future applications highly challenging in these systems.

The Fe-based superconductors provide an alternative pathway for pursuing Majorana modes at higher temperatures. $FeSe_xTe_{1-x}$ is the simplest compound in the Fe-based superconductor family with an optimum $T_c$=14.5K. This family of materials is highly attractive due to its versatility and tunability. Not only does this family of compounds grow well in thin film form[25], but its $T_c$ can be substantially enhanced through doping, pressure, and strain[26]. Through density functional theory it has been found that, for a range of concentrations around 50% Se, $FeSe_{0.5}Te_{0.5}$ possesses helical Dirac surface states due to band inversion along the Γ-Z direction[27-30]. In accordance with the Fu and Kane model[18], when an s-wave superconducting gap opens in the Dirac surface states (due to proximity to s-wave superconductivity in the bulk), it provides the ideal conditions for hosting MBS. In fact, there are multiple pieces of supportive evidence for this scenario in $FeSe_xTe_{1-x}$: high-resolution angle-resolved photoemission spectroscopy data reveal helical surface states that exhibit an s-wave gap below $T_c$, and a sharp zero-bias peak has already been observed inside vortex cores in this compound[31-33].

Interestingly, s-wave proximitized topological surface states can also host time-reversed pairs of dispersing 1D-Majorana states along domain walls (DWs) separating regions in which the superconducting order parameter is phase-shifted by π (ref.18). These modes possess a linear dispersion ($E = \pm v k_y$) with momentum parallel to the DW. This linear dispersion in 1D implies that these Majorana modes give rise to a constant density of states for energies below the superconducting gap, representing a clear signature of dispersing Majorana states that has not been observed to-date in any condensed matter system.

In this work, we use scanning tunneling microscopy (STM) to interrogate crystalline DWs in the proximitized Dirac surface states of $FeSe_{0.45}Te_{0.55}$ and search for signatures of 1D dispersing Majorana modes. We identify particular crystalline DWs that are associated with a half-unit cell shift of the Se atoms. Theoretical models specific to the FeSeTe family of compounds indicate that the superconducting order-parameter consistent with the crystal symmetries should develop a π-phase shift across this type of DW. Correspondingly, the STS spectra near the DWs show a remarkably constant density of states as a function of energy, a key signature of linearly dispersing 1D states that traverse the superconducting gap. Consistent with their topological nature, these DW states show a localization length that increases with increasing energy, eventually merging with the gap-edge continuum. Moreover, in stark contrast, all other kinds of defects that act purely as potential scatterers, e.g., step edges, impurities, and other 1D-defects, generate either a smaller superconducting gap, or sharp resonance peaks inside the gap. Taken together, our findings are consistent with the observation of helical 1D dispersing Majorana modes along DWs in $FeSe_{0.5}Te_{0.5}$.

Similar to most iron-based superconductors, the Fermi surface of $FeSe_{0.5}Te_{0.5}$ is composed of two hole-pockets (α' in red and β in green of Figure 1a) around the Γ-point, and two electron-pockets (γ in blue) at the BZ corner (M-point). According to theory, Te substitution into FeSe shifts the bulk $p_z$ band (found above $E_F$ in FeSe) downward towards the Fermi level[28]. This band then hybridizes with the $d_{xz}$ band (α band) to create a topological band inversion that pushes the α band about 14meV below $E_F$. In the resulting band gap, a topological Dirac surface state emerges, centered at the Γ-point on the (001) surface, as shown in Figure 1b. Below $T_c$, superconducting gaps are expected to open on both the surface and bulk bands[31].

In Figure 1c, we show an atomically resolved STM topographic image of the chalcogen (Se/Te) surface layer, where the Te atoms appear brighter than Se due to their more extended electronic orbitals. There is a marked absence of interstitial Fe atoms on the surface, usually observed as bright protrusions in the topography[34, 35]. This, in combination with the sharp superconducting transition (supplemental section 1, SS 1), confirms the high quality of these samples. We measure a series of differential conductance (dI/dV) spectra along the line shown in Fig. 1c and present them in Fig. 1d. Here all measurements were done at 0.3K. The data show that the spectral weight is completely suppressed to zero over a finite energy range ~ ±1 meV, and sharp peaks appear near the gap edge. These observations strongly suggest that there is no nodal structure in the gap function of $FeSe_{0.5}Te_{0.5}$, and the gap minima, if anisotropy exists, should be larger than 1meV.

There is an ongoing controversy regarding the gap values for each band reported by various ARPES and optical conductivity studies on similar materials[36-39]. STM is the ideal probe to measure gap values on different bands with high accuracy. However, due to the doped nature of FeSeTe, the gap values, as well as the number of gaps seen in any one spectrum, show spatial variations (Figure 1d). To obtain statistical information on the gap values and distribution, we record tunneling spectra dI(r, V)/dV on a densely spaced grid (240 x 240) over a 100x100nm field of view (FOV). The gap values are extracted through our multi-gap finding algorithm which finds the position of peak/shoulder features in each dI/dV spectrum, and accepts them as coherence peaks if they are particle-hole symmetric. A more detailed description of the algorithm and how we arrived at the gap map is presented in Methods. We classify the results by the number of gaps found for each spectrum and show a color-coded 2D map (gap map) in Figure 2a.

In general, we observe regions with one, two, or three gaps in the energy range of [-3.5meV, 3.5meV]. To visualize the evolution of the spectra as a function of position, a spectral line cut traversing the 3 regions (white line on the gap map) is shown in Figure 2b. One can see the spectra evolving from hosting two-gaps to three, and then to a single gap. The statistical analysis of the gap magnitudes performed by category (colored histograms), as well as the overall results, are shown in Figure 2c. For about 20% of the spectra taken in this FOV we can distinguish only one gap centered around 1.4meV; the two-gap spectra cover about 57% of this area, with mean gap values around 1.4 and 2.4meV; finally in the remaining area, three gaps can be detected simultaneously, with mean values at 1.4, 1.9 and 2.4meV. We note here that the largest gap value extracted from ARPES is around 4-5meV (ref.38). This larger gap is only seen as a hump-like feature in our spectra and was not picked up in our gap map due to the suppressed intensity of the peaks. However, this feature can be seen in line-cuts (SS 2). If we assign this hump-like feature (around ±4.5meV) to be the coherence peaks arising from the large superconducting gap on the γ sheet as shown in ref. 38, then, by comparison to ARPES, the peak features at 1.4 and 2.4 meV may be assigned to the smaller gaps on the α' and β bands, respectively. This suggests that, consistent with recent ARPES data, the 1.9 meV gap may be assigned to the topological surface state, affirming the topological nature of these samples[31].

Recent, STM measurements in Fe(Se,Te) have reported the existence of zero-bias conductance peaks inside vortex cores and near interstitial Fe atoms[32, 33, 35] which have been proposed to be signatures of zero-dimensional MBSs. We observe similar spectral line-shapes inside several vortex cores and near atomic scale defects (SS 3) in our samples, all of which are consistent with a topologically non-trivial surface state. Here we report the existence of one-dimensional dispersing Majorana modes near a new type of domain-wall defect. This defect is discovered by atomically-resolved topography as a 1D feature on the surface represented by a bright line (see Figure 3a). A zoomed-in view reveals that this bright line separates two crystal domains where the atoms on either side have a relative shift of half a unit cell in one direction (Fig. 3d). This shift is reflected as a split in reciprocal space Bragg peaks of the Fourier transform of the image (Fig. 3b and Fig.S5 in SS 4). The magnitude of the split in reciprocal space corresponds to a spatial scale of 12nm (half of the FOV of Figure 3a), which is consistent with the domain size in this FOV.

Differential conductance spectra obtained along three distinct paths traversing the DW (Fig. 3e) all reveal an intriguing evolution. As one approaches the DW, the superconducting coherence peaks in the dI/dV spectra are suppressed and new electronic states emerge inside the gap, resulting in a near featureless, flat dI/dV at the domain wall center as shown by the highlighted lines in Fig. 3e and the inset (also see Fig. S6 in SS 4).

An explanation of this observation is the existence of linearly dispersing Majorana states at the DW, as it would naturally give rise to a constant dI/dV in 1D. According to the Fu-Kane model[18], realizing 1D dispersing Majorana states would require three key ingredients: non-trivial topological surface states, s-wave superconductivity that gaps the surface states, and a π-phase shift in the superconducting order parameter across the domain wall[18]. Our detailed gap-maps already indicate the presence of proximity-gapped Dirac surface states, thus satisfying the first two criteria. This leaves us with the question of how to generate a superconducting phase shift. An s± order parameter may in general host a π-phase shift across the domain wall. Moreover, theoretical studies of iron-based superconductors which consider the full lattice and orbital symmetries, suggest that they can host an s-wave spin-singlet pairing state where our type of domain wall, having a half unit-cell shift, can naturally give rise to a π-phase shift in the superconducting order parameter[40] as depicted in Figure 3e. Thus, this system has all the essential ingredients necessary for hosting dispersing Majorana modes.

Further support for the identification of the observed domain wall modes as Majorana modes comes from the study of other types of extended 1D defects (Fig. S8 and S9 in SS 4) that do not intrinsically give rise to a π-phase shift, e.g., step edges. Indeed, our experimental dI/dV data reveal that step edges on the surface of FeSe$_{0.45}$Te$_{0.55}$ induce bound states inside the superconducting gap, but do not give rise to a constant DOS, since one important requirement for the emergence of Majorana mode, the π-phase shift, is absent.

One might wonder whether the experimentally observed DW modes could also possess a topologically trivial origin, unrelated to existence of a topological surface state. Based on previous studies, the superconducting order parameter in Fe(Se,Te) is expected to be a sign-changing s± state[41, 42]. In such a state, defects, regardless of their magnetic properties, would induce impurity states inside the superconducting gap. The experimentally observed domain wall – representing a 1D defect could therefore lead to the emergence of an impurity band inside the superconducting gap even in a topologically trivial phase. To investigate this possibility, we employ a theoretical model for a topologically trivial superconducting state of FeSe$_x$Te$_{1-x}$ (see Ref. 43) and represent the DW as a line of potential scatterers (SI Sec.5). We find, as expected, that the DW gives rise to an impurity states inside the superconducting gap. However, these states do not in general traverse the superconducting gap (only for fine-tuned values of the scattering potential do impurity states near zero energy emerge). Moreover, such states are not uniformly distributed in energy inside the gap and cannot result in the observed constant density of states. The same conclusion also holds if the DW-separated, π-phase shifted superconducting regions are present in an otherwise topologically trivial phase (Fig. S11 in SS 5).

These findings are further confirmed by our experimental study of twin-domain walls in the topologically trivial but related superconducting compound, FeSe (Fig. S12 in SS 6). While such domain walls give rise to a suppression of the superconducting gap, they do not result in a constant DOS. These theoretical and experimental findings taken together make it very unlikely that the observed constant density of states near the DW can arise in a topologically trivial superconducting phase, which further emphasizes the important role played by non-trivial topology in creating the observed domain wall modes.

The topological nature of the DW modes necessarily dictates a specific spatial and energy dependence of the DW states. In particular, with increasing energy, the Majorana mode must continuously evolve from being localized at the DW at zero energy to being delocalized as their energy reaches the gap edge. In other words, the modes' localization length increases with increasing energy, leading to an increase in the modes' spatial extent from the DW. To visualize this evolution, we obtained spatial differential conductance maps (dI/dV maps) in the vicinity of the DW ranging in energies from $E_F$ up to 1.4 mV where the first set of coherent peaks is located (Figs. 4a-f). At $E_F$, the in-gap states are confined within an approximately 3nm width. These states begin to expand in real space with increasing energy, and become less visible at 0.85mV due to a lack of contrast in the intensity with respect to the rest of the area. The dI/dV maps show that the domain wall states are present at all energies inside the gap. However, the spatial extent of the states grows with increasing energy consistent with a topological origin for these modes.

FeSe$_{0.5}$Te$_{0.5}$ may have provided the first glimpse into linearly dispersing 1D Majorana modes. This immediately opens the door to new experiments if one can locate the DWs on the surface and fabricate contacts or other hetero-layers that couple to the DW. For example, one could try to measure the fractional (4π) Josephson effect using a flux loop, or deposit magnetic layers to generate chiral Majorana modes or even MBS. Beyond the Majorana fermion context, our experimental results have two important implications. First, our observations provide supporting evidence for the existence of topological surface states and a Fu-Kane proximitized superconducting state in $FeSe_{0.5}Te_{0.5}$. Second, the connection between the crystal domain wall and the superconducting π-phase shift provides evidence in support of a superconducting order parameter with a real-space sign-change within a unit cell[40, 42].

## Methods

Large single crystals of $FeSe_{0.45}Te_{0.55}$ were grown using the self-flux method, and the superconducting transition temperature is about 14.5K. Fig. S1 shows the temperature dependence of the mass magnetization. From previous studies of $FeSe_xTe_{1-x}$, sample quality is known to be important for the observation of sharp spectra[34], as well as MBS in vortex cores in the topological phase regime[32]. In particular, the presence of Fe interstitials can weaken superconductivity. For this reason, we chose the highest quality samples which did not require oxygen annealing for the emergence of superconductivity. All the data shown in this work are taken on the as-grown samples.

We used chemically etched tungsten tips annealed to an orange color in UHV. Before performing measurements on $FeSe_{0.45}Te_{0.55}$, we check the tip quality on the surface of a single crystal Cu(111). The single crystals $FeSe_{0.45}Te_{0.55}$ were cleaved *in situ* at 90K and immediately insert into STM head. All STM data were acquired at 0.3K. Spectroscopic measurements were taken by a standard lock-in technique at a frequency of 987.5Hz, under modulation ~0.06meV.

The gap algorithm mentioned in the main text reads the spectra taken at each pixel and applies a standard Gaussian filter around a two-pixel radius in real space. Then, employing the finite difference method, it calculates the derivative of each spectrum measured. Both the discrete spectrum and its derivative are interpolated using third degree polynomials. Mathematica's build in root finding algorithm is then used to find the location of the zeros in the interpolated derivative, and then discriminated against the value of its second derivative used to ensure it is a local maximum. The accepted values are categorized as peaks. To find shoulder like features, the algorithm looks for saddle-points instead. The peak and shoulder features are further filtered by using a small energy window to ensure they are particle hole symmetric, where a narrower filter is used for the shoulder features. The gap values at each pixel are calculated by taking the average value of the energy positions at which the accepted features are found. Finally, each pixel is color coded by the number of gaps in it and the map reassembled.

## Acknowledgements

We thank Hong Ding, Steve Kivelson, and Charlie Kane for useful conversations. The work in Brookhaven is supported by the Office of Science, U.S. Department of Energy under Contract No. DE-SC0012704   V.M. gratefully acknowledges partial support from NSF Award No. DMR-1610143 and the US Department of Energy under Award Number DE-SC0014335 for the STM studies. T.L.H. thanks the US National Science Foundation under the MRSEC program under NSF Award Number DMR-1720633 (SuperSEED) for support. M.G. and D.K.M. acknowledge support from the U. S. Department of Energy, Office of Science, Basic Energy Sciences, under Award No. DE-FG02-05ER46225.

## Competing financial interests

The authors declare no competing interests.

## Corresponding author

Correspondence to: Vidya Madhavan(vm1@illinois.edu)

# Figure 1

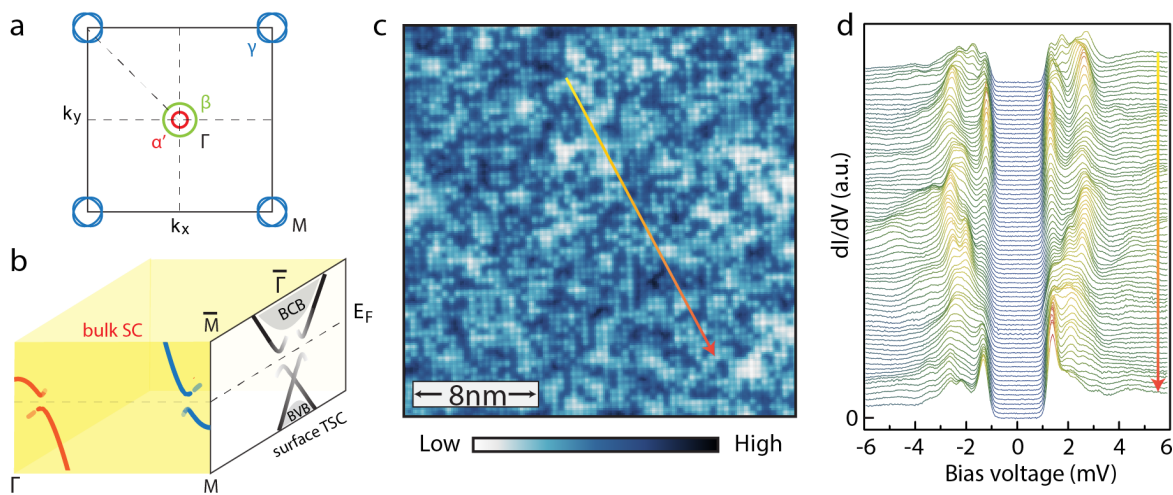

**Figure 1. Band structure and superconductivity in FeSe$_{0.45}$Te$_{0.55}$. a,** Sketch of Bulk Fermi surfaces of Fe(Se, Te) at $k_z$=0. **b,** Cartoon image showing superconductivity in the bulk and proximitized superconductivity in the topological surface state[31]. **c,** Topographic image in a 25nm X 25nm field of view (bias voltage $V_S$=40mV, tunneling current $I_t$=100pA). **d,** STS spectra taken along the line shown in c.

**Figure 2**

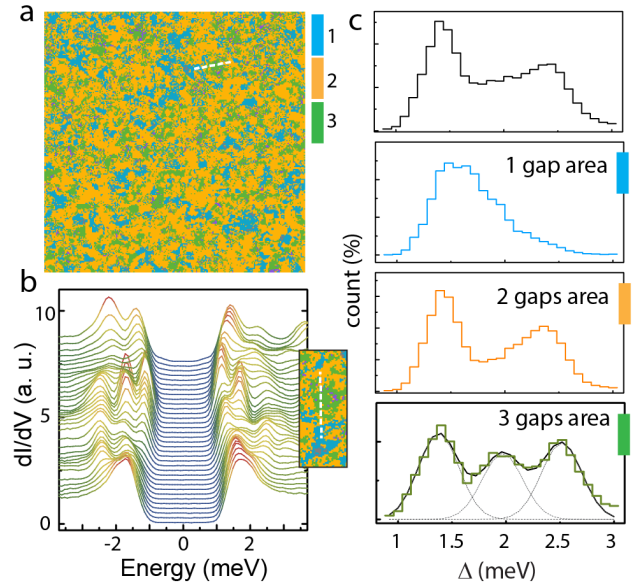

**Figure 2. Statistical analysis of superconducting gaps**. **a**, 100nm X 100nm map depicting the distribution of superconducting gaps in FeSe$_{0.45}$Te$_{0.55}$. Blue, orange, and green colors indicate whether a single gap, two gaps, or three gaps were found at each pixel. To construct the map, dI/dV spectra we obtained on a 240 X 240 pixel grid. The gap values at each pixel were obtained through a multi peak finding algorithm. **b**. STS spectra taken along the white line shown in a, starting with a two-gap region (orange) which transitions to a three-gap region (green), finally ending in a single- gap region (blue). **c,** Histogram of the gap values in the one, two and three gap regions. The dark curves show Gaussian fit to the gap distribution, with mean gap values of 1.4, 1.9 and 2.4 meV.

# Figure 3

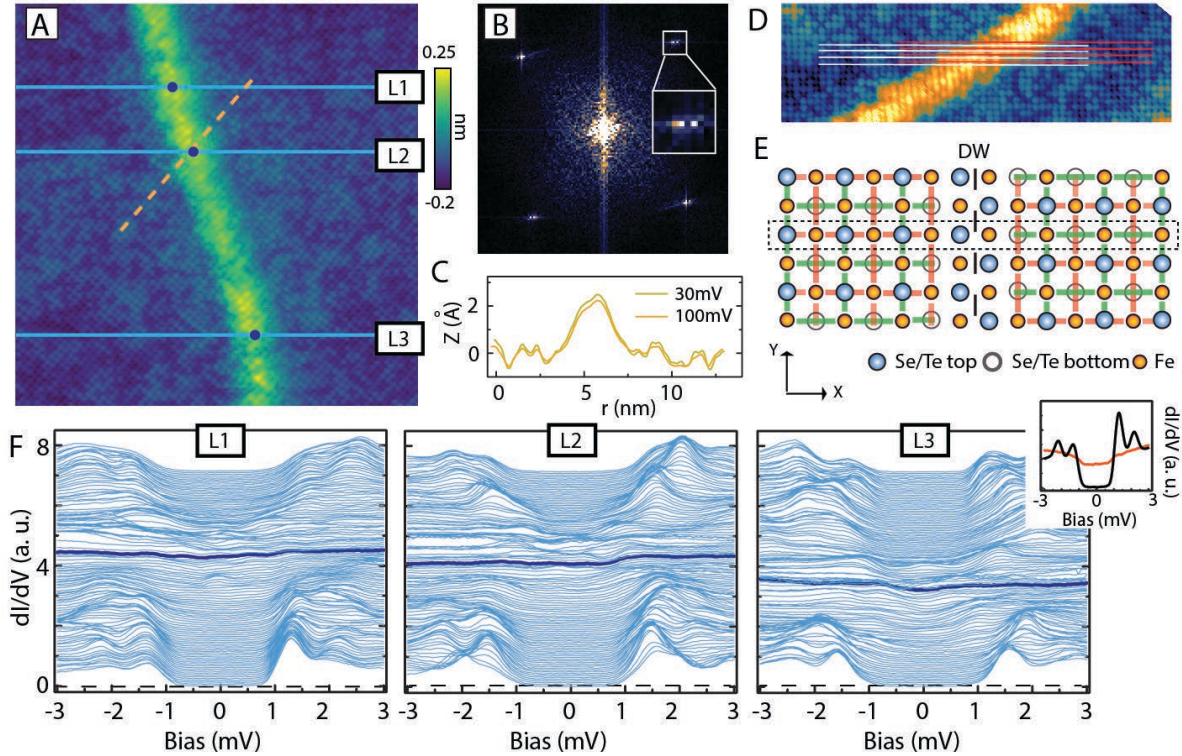

**Figure 3. Signature of dispersing 1D Majorana modes on a domain wall (DW)**. **a**. 25nm X 25nm topographic image showing a DW. **b**. 2D FFT of **a**, showing the splitting of the Bragg peaks which indicates the presence of domains in this image. Inset: zoom-in near one of the Bragg peaks. **c**. Height scans taken at different bias voltage along the yellow dashed line in **a**. **d**. Zoom in of the DW. The dashed lines track the atomic lattice on both sides of the DW. A half-unit cell shift can be observed between one side and the other. **e**. Schematic of the half unit cell shift across the DW. The schematic also depicts how one might obtain a π-phase in the superconducting order parameter shift across such a DW. Superimposed on the lattice are red and green bars, which denote the parity of next-nearest neighbor pairing (ref. 40 and 42). Concentrating now on the atoms inside the dashed box, one observes that the parity shifts from red on the left of the DW, to green on the right. This creates a π-phase shift in the superconducting order parameter. **f.** Linecut profiles of dI/dV spectra along the three blue lines in **a** (L1, L2 and L3 shown in sequence from left to right), which cross the DW. The spectra shape obtained right on the DW (at position of dots in **a**) are highlighted with a dark blue color. A direct comparison of the spectra taken on the DW (orange) and far away (dark) is shown in the inset.

# Figure 4

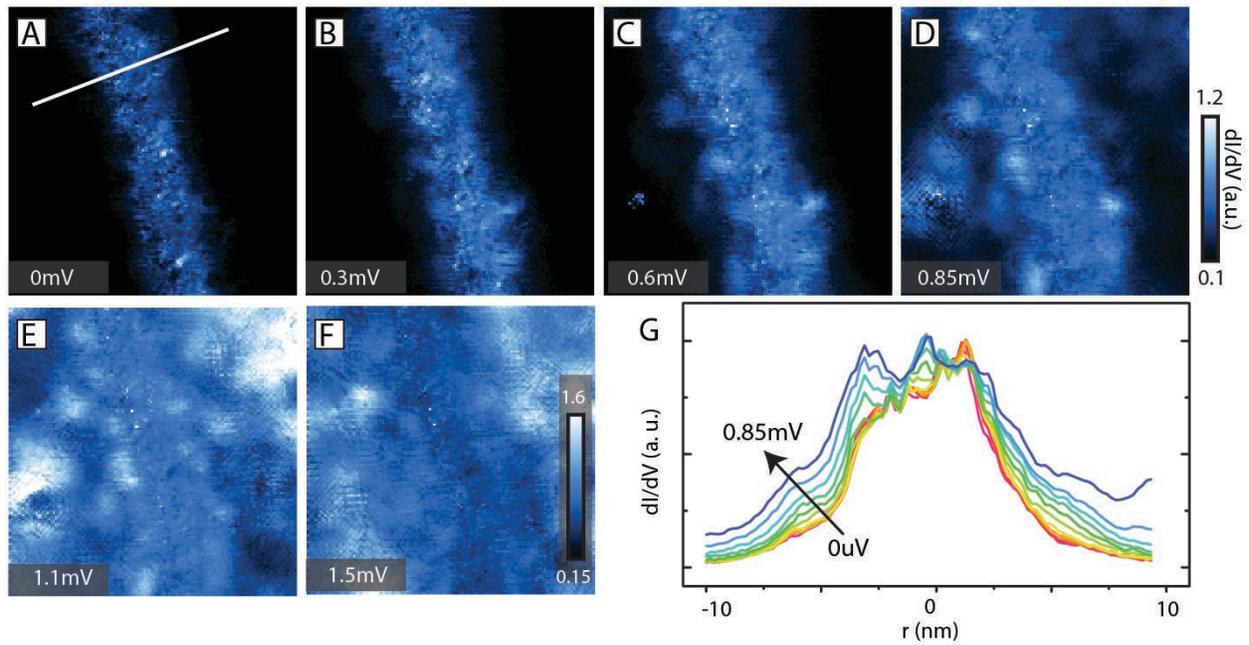

**Figure 4. Spatial distribution of the 1D Majorana mode at a DW with increasing energy**. **a**-**f**. dI/dV map from 0-1.5meV. The maps are 25x25nm in size and spectra were obtained on a 130x130 pixel grid. Domain wall states are present at all energies inside the gap upto 1meV when the states merge into the coherence peaks. However, the spatial extent of the states grows with increasing energy. **g**. LDOS profiles measured at different energies along the white line perpendicular to the DW.

# Supplementary information for
# "Signature of Dispersing 1D Majorana Channels in an Iron-based Superconductor"

Zhenyu Wang, Jorge Olivares Rodriguez, Martin Graham, Genda Gu, Taylor Hughes, Dirk K. Morr, and Vidya Madhavan

**Supplemental Section 1**

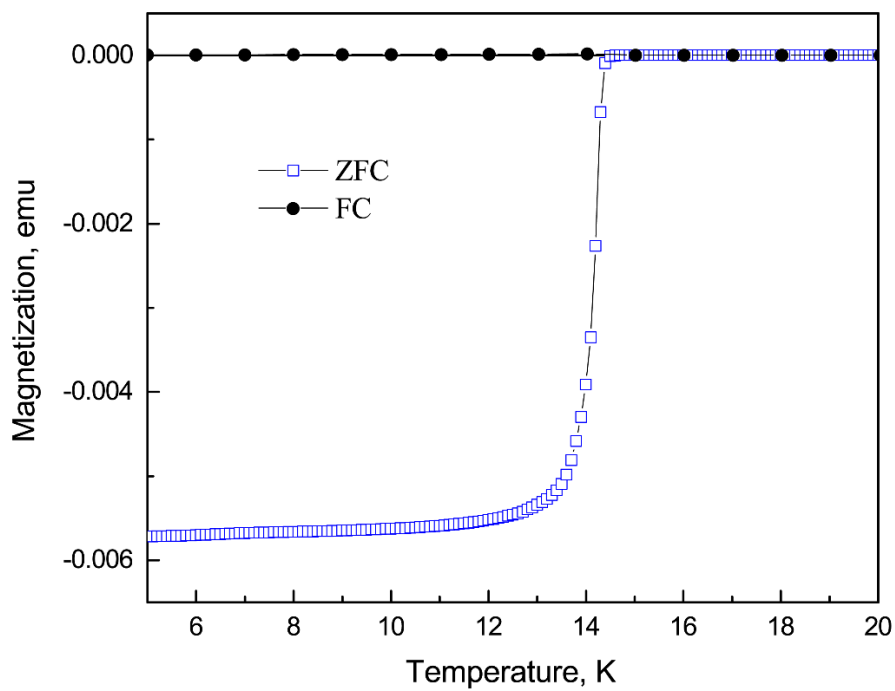

**Figure S1. Superconducting transition.** Temperature dependence of mass magnetization after zero-field cooled (ZFC) and field-cooled (FC) modes at 5 Oe.

**Supplemental Section 2**

We extract the gap values in 150 spectra through our multi-gap finding algorithm and plot the histogram in Fig. S2a. Other than peaks at 1.4, 2.1 and 2.5meV, we can see particle-hole symmetric peaks at higher energy, between 4 and 5meV. A similar 4-5meV gap has also been reported in ARPES and optical measurements [1-3]. Fig.S2b shows a comparison of one of our STM spectra and ARPES data [1, 4].

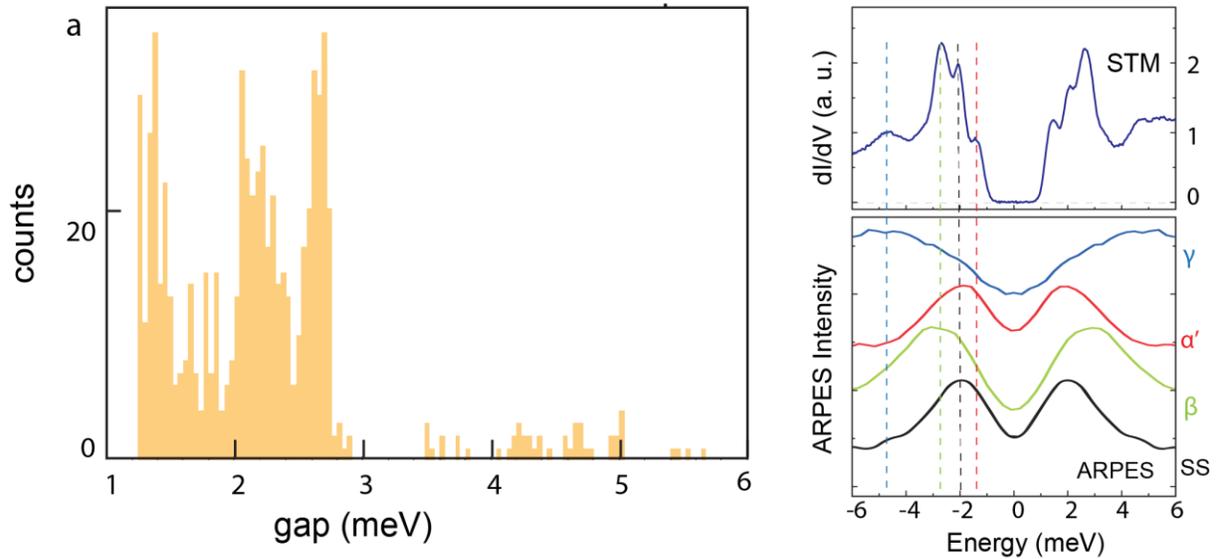

**Figure S2.** (a) Histogram of the particle-hole symmetric peaks in the line cut. (b) Comparison between STS and ARPES data. ARPES data are reproduced from [1, 4].

## Supplemental Section 3

Recent STM/STS measurements have revealed zero-energy bound states inside the vortex cores and at interstitial iron impurities in Fe(Se,Te), which are suggested to be Majorana bound states [5-7]. We have performed similar measurements on our samples.

We first image the vortices at E=0meV with a 3T magnetic field perpendicular to the sample surface (Fig. S3a). Detailed tunneling spectra near the vortex core are shown in Fig. S3b, labeled as #1, #2 and #3, respectively. For all the vortex cores states, we see a suppression of the spectral weight near the superconducting gaps and clear bounds states inside the gap. However, the energy position of these bound states vary. For example, linecuts #1 and #3 show zero-bias peaks, whereas #2 does not.

We then perform spectroscopic measurements in zero magnetic field to characterize the impurity effects. In Fig. S4, we show three kinds of impurities and the local electronic states around them. Near impurity 1 (located at top Se/Te sites), we see a pronounced zero-energy bound state around it. However, the peak location is spatially dependent, and moves away from zero energy right on the impurity. Most of the impurities show particle- hole symmetric bound states inside the gap, like impurity 2. We also find non-split zero-energy peaks near the interstitial iron atoms as reported in [7].

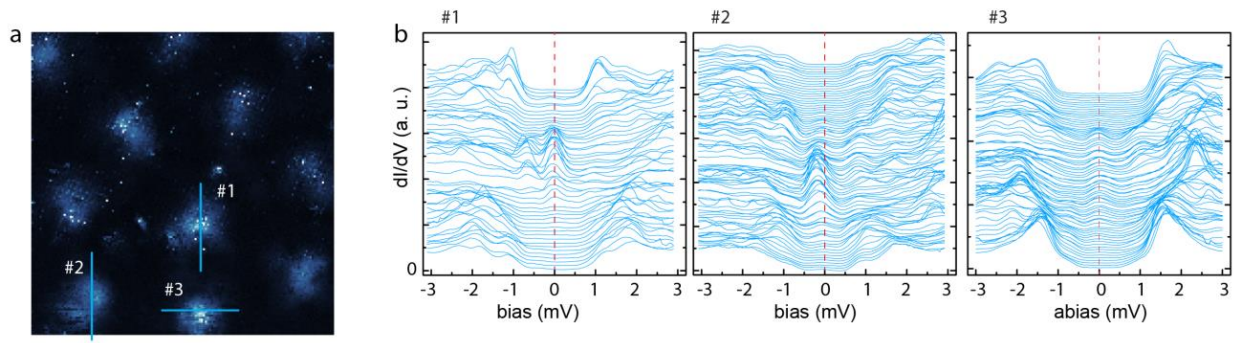

**Figure S3. Zero-bias peaks inside the vortex core.** (a) A ZBC map to show the vortex cores. (b) A waterfall-like plot of tunneling spectra measured along the cyan lines in a. ZBP can be found in some of the vortex cores (#1 and #3). Data was taken with 3T magnetic field perpendicular to the sample surface.

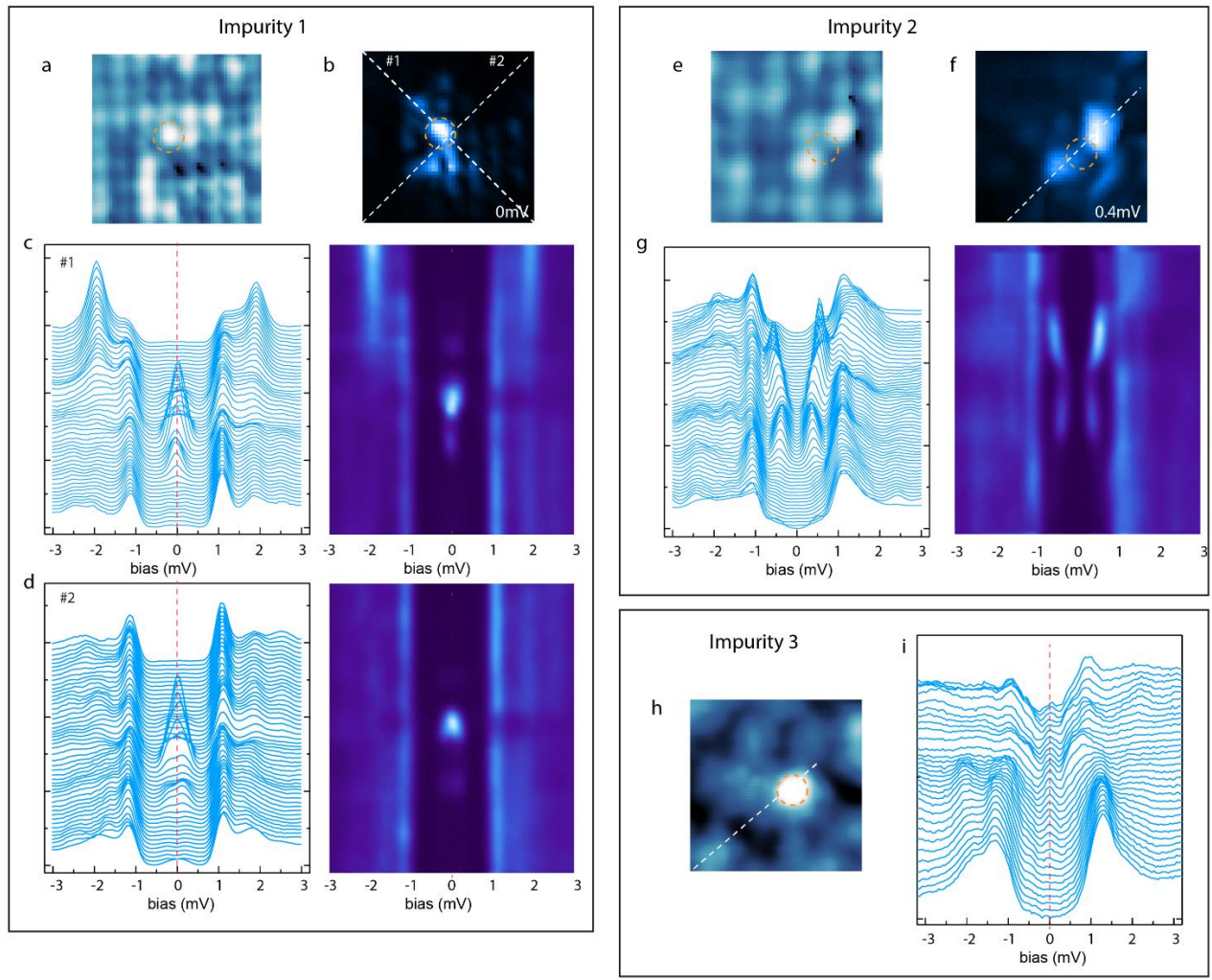

**Figure S4. Bound states near individual impurities.** All three kinds of impurity show clear bound states inside the gap. Zero-energy bound states can be seen near the interstitial iron atoms.

**Supplemental Section 4**

Our experiments have revealed a constant DOS inside the superconducting gap on the DWs where the crystalline lattice on either side have a relative shift of half a unit cell. We show a detailed FFT analysis in figure S5.

This lattice shift is critical to generate a phase difference of the superconducting order parameter in Fe(Se, Te), and then leads to the emergence of 1D Majorana channels. Linecuts towards and along the domain wall are shown in Fig. S6, and a zoom-in around the domain is shown in Fig. S7.

We have measured the DOS at other kinds of 1D defects. In Fig. S8 we present two kinds of extended 1D defects which show a similar height ~ 0.2nm in the topography. However, we do not see any Bragg peak splitting in the FFT, indicating the absence of a lattice shift across these 1D defects. At the center of these defects, we continue to observe superconducting gap features which is in strong contrast with the flat DOS near the DW.

In Fig. S9, we present the tunneling spectra cross a step edge. One can see that the step edge creates pronounced bound states inside gap. All these findings indicate that the constant DOS observed near the DW is quite unique and cannot be solely explained by potential scattering or structural distortions.

**Characterizing the DW**

Fourier transforms of areas with no domain walls show sharp Bragg peaks in the FFT as shown in a. When DWs are included in the field of view, Bragg peaks split into two peaks, as shown in b and d. We have carried out extensive analysis to confirm that this splitting is actually induced by the lattice shift between the two domains. First, we mask the domain wall before taking the FT. The corresponding FFTs shown in c and e continue to show the splitting indicating that the Domain wall itself is not responsible for the splitting. Second, upon masking one of the two domains the peak splitting goes away i.e., we find a single Bragg peak in the FFT (f and g). Finally, the magnitude of the splitting is supposed to represent the domain size. We confirm this for the two images shown in Fig. b and d where the domains occupy approximately half the image in each. The magnitude of the splitting is 1/31 and 1/50 of the Bragg peak value for Fig. b and d, respectively. These correspond to 12nm and 19nm in real space which is half of the image size.

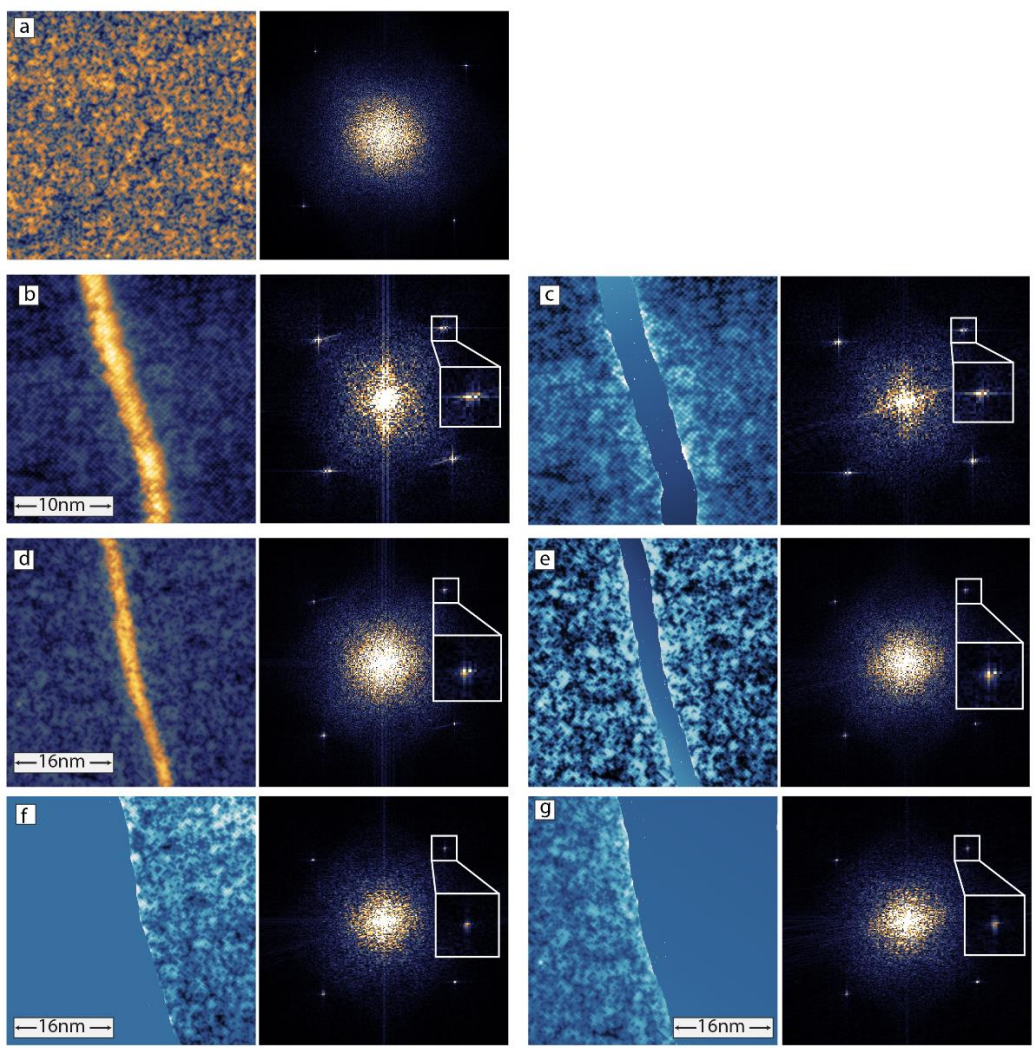

**Figure S5. Characterizing the DW.** STM images and their corresponding FFT are shown.

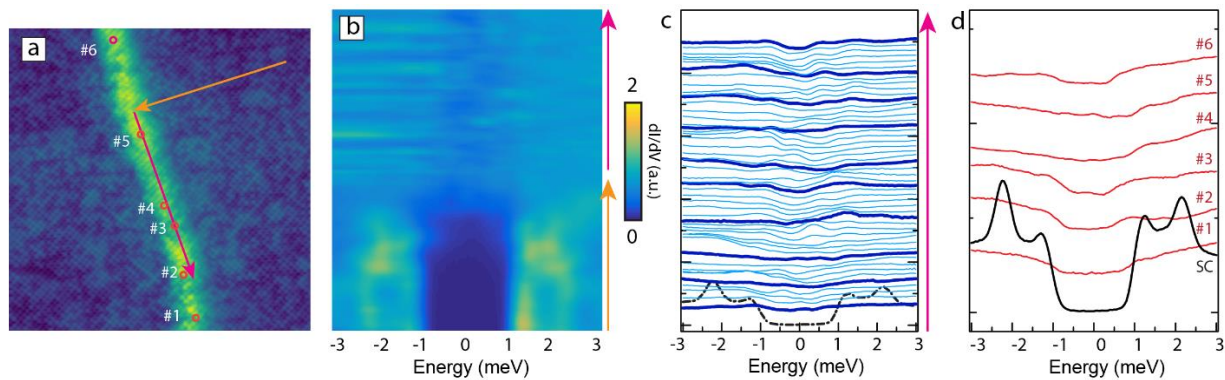

**Figure S6. Measuring the DOS across and on the DW.** (b) Spectra along a line (linecut) shown as an intensity plot. The path over which spectra were obtained first goes perpendicular to the DW (as indicated by orange arrow) and the along (as shown by red arrow) the center of the DW. (c) Same linecut along the red line at the center of the domain wall reiterating that the spectra host a flat density of states. Figure d shows a few individual spectra in greater detail. These were obtained at the positions denoted by the circles marked in (a). A superconducting spectrum away from the domain wall is also shown for comparison.

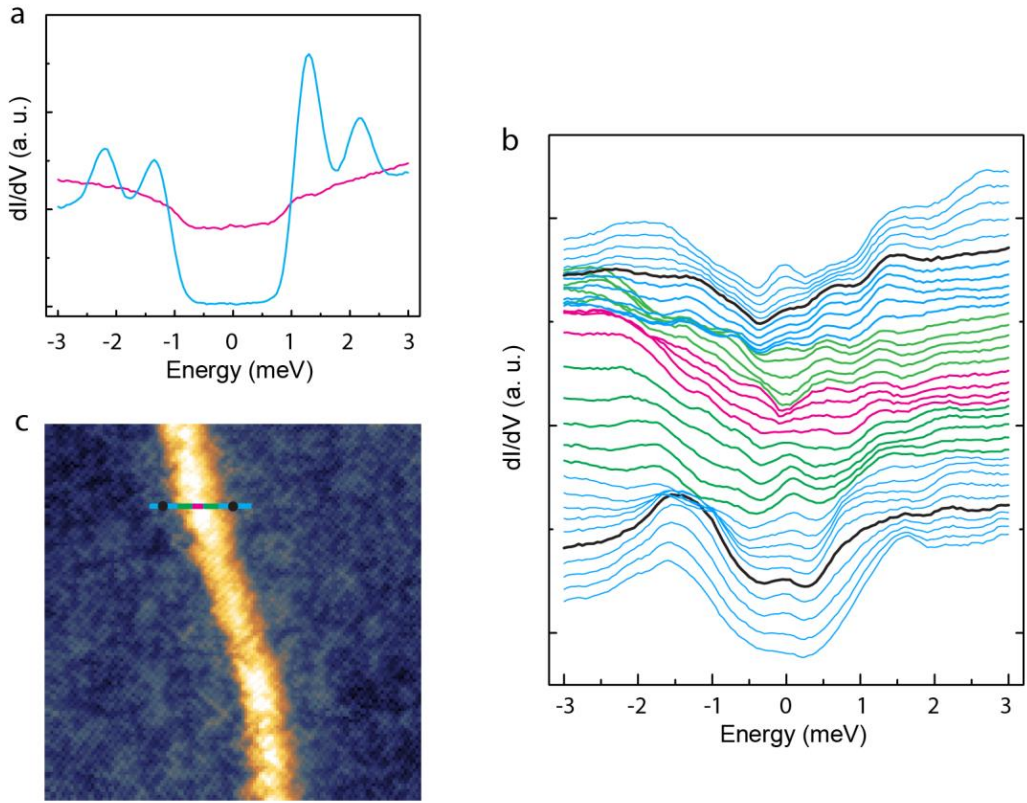

**Figure S7. Spectra measured near the DW.** (a) The spectrum at the center of the DW (pink) compared to superconducting spectrum far from DW. (b) Line across the DW where the spectra shown in (c) were obtained. (c) Detailed spatial dependence of shape of spectra across the DW. Spectra shown were obtained about 0.19nm apart.

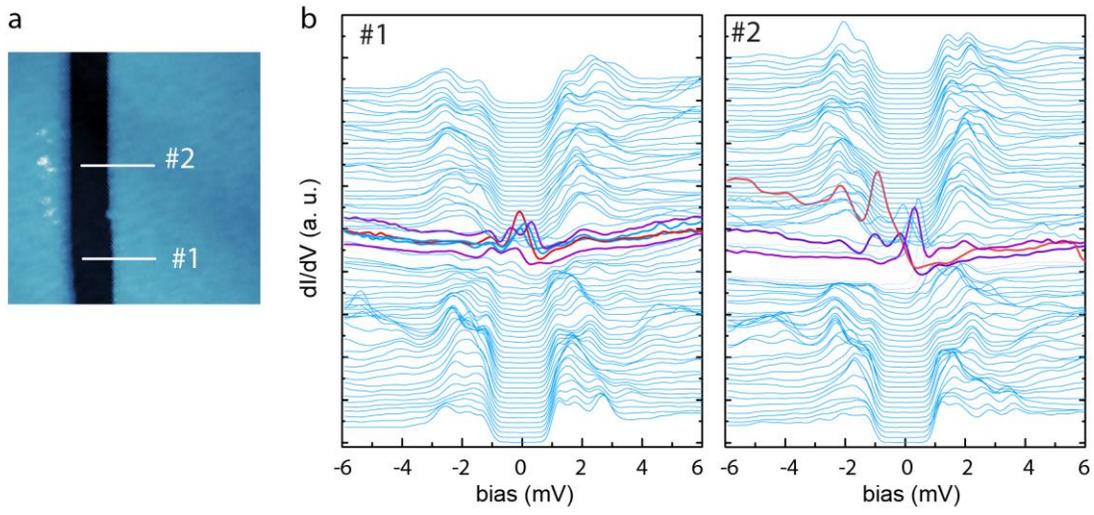

**Figure S8. Step edge.** We observe pronounced bound states inside the superconducting gap near a step edge.

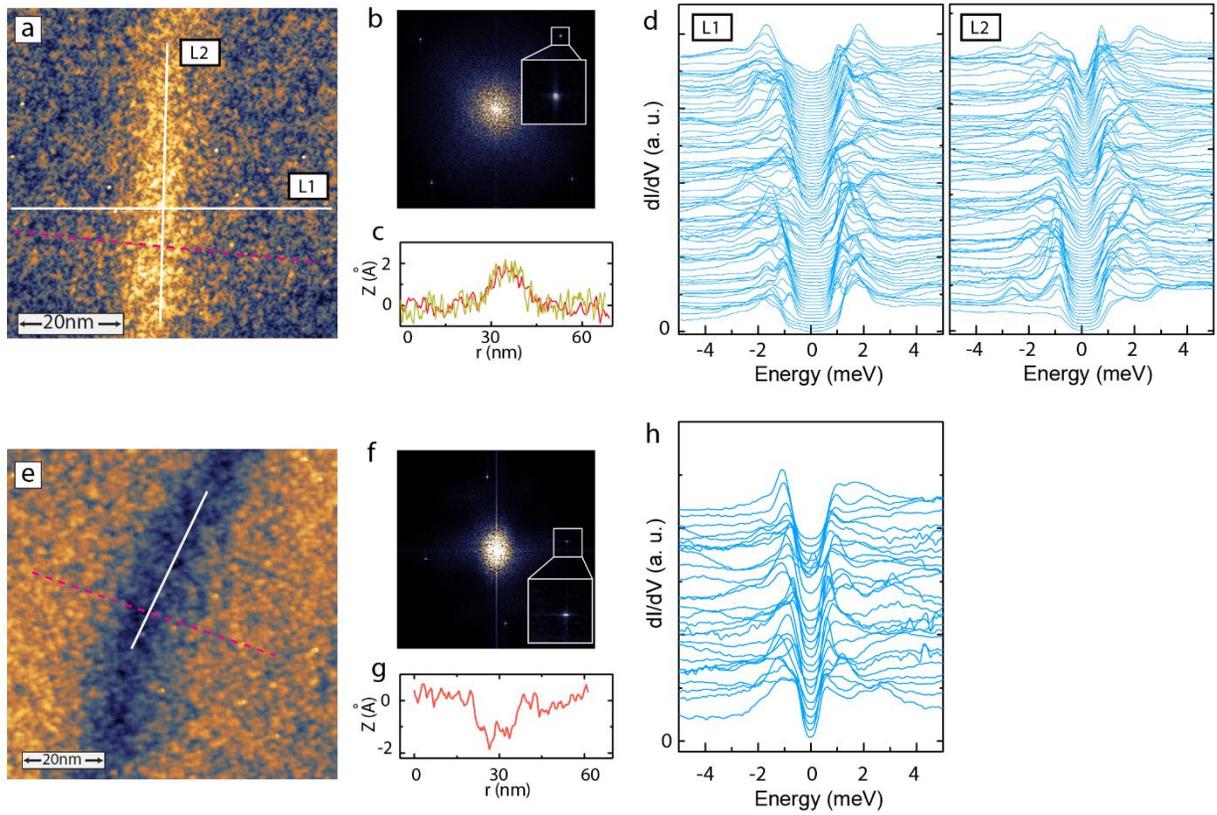

**Figure S9. Other kinds of 1d defects.** (a) and (e) show other two kinds of 1D defects. No lattice shift has been found near these two kinds of 1D defects in the FFT (b and f). We observe superconducting features on these defects (d and h), even though the height of these defects are similar to the DW we shown in the main text (c and g).

## Supplemental Section 5

To describe the electronic structure of Fe(Se, Te) we employ a 5-band tight-binding model [8] that was previously employed to describe the differential conductance measured in STS experiments. In real space, the Hamiltonian is given by

$$H_0 = \sum_{r,\sigma} \sum_{\alpha=1}^{5} \varepsilon_{\alpha\alpha} c^\dagger_{r,\alpha,\sigma} c_{r,\alpha,\sigma} - \sum_{r,r',\sigma} \sum_{\alpha,\beta=1}^{5} t^{\alpha,\beta}_{r,r'} c^\dagger_{r,\alpha,\sigma} c_{r',\beta,\sigma}$$

$$H^{MF}_{SC} = \sum_{<r,r'>} \sum_{\alpha=1}^{5} \Delta_{\alpha\alpha} c^\dagger_{r,\alpha,\uparrow} c^\dagger_{r',\alpha,\downarrow} + H.c.$$

Where $\alpha, \beta = 1, ...,5$ are the orbital indices corresponding to the $d_{xz}, d_{yz}, d_{x^2-y^2}, d_{xy}$, and $d_{3z^2-r^2}$ orbitals, respectively, $-t^{\alpha,\beta}_{r,r'}$ represents the electronic hopping amplitude between orbital α at site **r** and orbital β at site **r'**, $\varepsilon_{\alpha\alpha}$ is the on-site energy of orbital α. $-t^{\alpha,\beta}_{r,r'}$ is non-zero to the fifth-nearest neighbor, where $|r - r'| = 3$. In momentum space, the hopping parameters lead to the following intra- and inter-orbital dispersions

$$\varepsilon_{11/22}(k) = 2t^{11}_{x/y} cos k_x + 2t^{11}_{y/x} cos k_y + 4t^{11}_{xy} cos k_x cos k_y \pm 2t^{11}_{xx} (cos2k_x - cos2k_y)$$
$$+ 4t^{11}_{xxy/xyy} cos2k_x cos k_y + 4t^{11}_{xyy/xxy} cos2k_y cos k_x + 4t^{11}_{xxyy} cos2k_x cos2k_y$$

$$\varepsilon_{33}(k) = 2t^{33}_{x}(cos k_x + cos k_y) + 4t^{33}_{xy} cos k_x cos k_y + 2t^{33}_{xx}(cos2k_x + cos2k_y)$$

$$\varepsilon_{44}(k) = 2t^{44}_{x}(cos k_y + cos k_y) + 4t^{44}_{xy} cos k_x cos k_y + 2t^{44}_{xx}(cos2k_x + cos2k_y)$$
$$+ 4t^{44}_{xxy}(cos2k_x cos k_y + cos2k_y cos k_y) + 4t^{44}_{xxyy} cos2k_x cos2k_y$$

$$\varepsilon_{55}(k) = 2t^{55}_{x}(cos k_y + cos k_y) + 2t^{55}_{xx}(cos2k_x + cos2k_y)$$
$$+ 4t^{55}_{xxy}(cos2k_x cos k_y + cos2k_y cos k_y) + 4t^{55}_{xxyy} cos2k_x cos2k_y$$

$$\varepsilon_{12}(k) = -4t^{12}_{xy} sin k_x sin k_y - 4t^{12}_{xxy}(sin2k_x sin k_y + sin2k_y sin k_x) - 4t^{12}_{xxyy} sin2k_x sin2k_y$$

$$\varepsilon_{13/23}(k) = \pm 4it^{13}_{x} sin k_{y/x} \pm 4it^{13}_{xy} sin k_{y/x} cos k_{x/y} \mp 4it^{13}_{xxy}(sin2k_{y/x} cos k_{x/y} - cos2k_{x/y} sin k_{y/x})$$

$$\varepsilon_{14/24}(k) = 2it^{14}_{x} sin k_{x/y} + 4it^{14}_{xy} cos k_{y/x} sin k_{x/y} + 4it^{14}_{xxy} sin2k_{x/y} cos k_{y/x}$$

$$\varepsilon_{15/25}(k) = 2it^{15}_{x} sin k_{y/x} - 4it^{15}_{xy} sin k_{y/x} cos k_{x/y} - 4it^{15}_{xxyy} sin2k_{y/x} cos2k_{x/y}$$

$$\varepsilon_{34}(k) = 4t^{34}_{xy}(sin2k_y sin k_x - sin2k_x sin k_y)$$

$$\varepsilon_{35}(k) = 2t^{35}_{x}(cos k_x - cos k_y) + 4t^{35}_{xxy}(cos2k_x cos k_y - cos2k_y cos k_x)$$

$$\varepsilon_{45}(k) = 4t_{xy}^{45}\sin k_x \sin k_y + 4t_{xxyy}^{45}\sin 2k_x \sin 2k_y)$$

The on-site energies for 5 orbitals are given by (7.0, 7.0,-25, 20, -25.1meV), while the hopping are given in the tables (in meV).

|      | i=x   | i=y   | i=xy  | i=xx | i=xxy | i=xyy | i=xxyy |
|------|-------|-------|-------|------|-------|-------|--------|
| m=1  | -11.0 | -43.0 | 28.0  | 2.0  | -3.5  | 0.5   | 3.5    |
| m=3  | 32.0  |       | -10.5 | -2.0 |       |       |        |
| m=4  | 22.0  |       | 15.0  | -2.0 | -3.0  |       | -3.0   |
| m=5  | -10.0 |       |       | -4.0 | 2.0   |       | -1.0   |

Table 1. Intra-orbital hopping (meV)

|       | i=x   | i=xy  | i=xxy | i=xxyy |
|-------|-------|-------|-------|--------|
| mm=12 |       | 5.0   | -1.5  | 3.5    |
| mm=13 | -35.4 | 9.9   | 2.1   |        |
| mm=14 | 33.9  | 1.4   | 2.8   |        |
| mm=15 | -19.8 | -8.5  |       | -1.4   |
| mm=34 |       |       | -1.0  |        |
| mm=35 | -30.0 |       | -5.0  |        |
| mm=45 |       | -15.0 |       | 1.0    |

Table 2. Inter-orbital hopping (meV)

To match the experimental spectra, we use nonzero superconducting order parameters in three orbitals: $\Delta_{xz} = \Delta_{yz} = 0.55 meV$, $\Delta_{xy} = 0.38 meV$ and $\Delta_{\alpha\beta} = 0$ in all other orbitals. Assuming pairing between next-nearest Fe sites then yields an s+- wave states. The resulting density of states well describes the experimental data, as shown in Fig. S10b.

To model a line defect, we introduce an on-site non-magnetic scattering defects given by

$$H_{scat} = \sum_{R,\sigma} \sum_{\alpha=1}^{5} U_0 c_{R,\alpha,\sigma}^\dagger c_{R,\alpha,\sigma}$$

Where the sum runs over all defect sites in a line along the crystal axis. The total Hamiltonian in real space in the superconducting states is then given by $H = H_0 + H_{SC}^{MF} + H_{scat}$. Using Nambu spinors $\Psi^\dagger$, $\Psi$, the Hamiltonian can be written as

$$H = \Psi^\dagger \hat{H} \Psi$$

Diagonalization of $\hat{H}$ then yields the eigenenergies of the system. The local, orbitally-summed density of states is then obtained from

$$N(\mathbf{r}, E) = -\frac{1}{\pi} \sum_\alpha \mathrm{Im} G_{\alpha,\alpha}(\mathbf{r},\mathbf{r}, E)$$

Where $G_{\alpha,\alpha}(\mathbf{r},\mathbf{r}, E)$ is the local retarded Green function of orbital α. These are given as the diagonal elements of the Greens function matrix

$$\hat{G}^{-1} = (\omega + i\delta)\hat{1} - \hat{H}$$

In Fig. S10a we present the lowest eigenergies of the system as a function of the defect lines scattering strength, U. For comparison we present the DOS of the clean systems in b. With increasing U, impurity states are pulled into the gap, and cross zero energy around U~60meV and 90meV. However, the defect line does give rise to generic zero energy states.

Next, we consider the effects of a linear domain wall in the superconductor that separates two regions in which the SC order parameters (SCOPs) possess a π-phase shift, and hence differ in their overall sign. In this case, the SCOP is position-dependent

$$\mathrm{H}_{SC}^{MF} = \sum_{\langle r,r'\rangle} \sum_{\alpha=1}^{5} \Delta_{\alpha\alpha} c^\dagger_{r,\alpha,\uparrow} c^\dagger_{r',\alpha,\downarrow} + H.c.$$

In Fig. S11a, we present the local density of states near the π-phase domain wall. While the domain wall clearly leads to the emergence of in-gap states, these in-gap states are located at finite energy (we do not find that these energy states move to lower energies with increasing system size). Thus, a π-phase domain wall does not lead to the emergence of generic zero-energy states, even though states may be located close to zero energy.

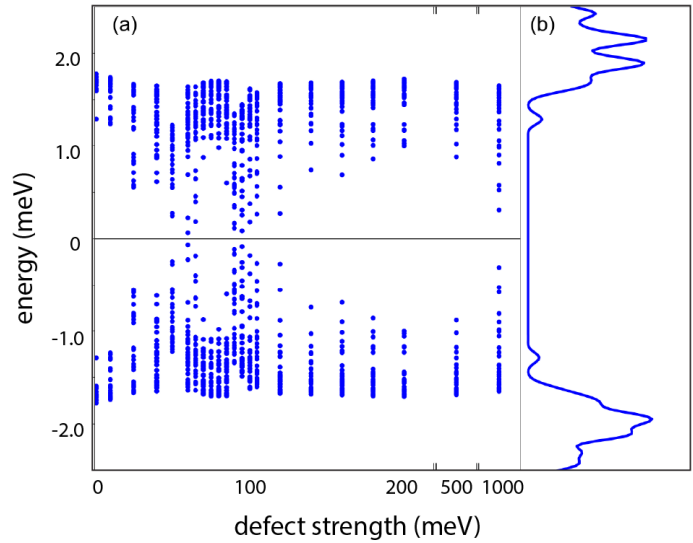

**Figure S10. Theoretical simulation of potential scattering due to a line defect**. (a) Lowest-energy electronic states for a 121x121-site system, as a function of the defect line's scattering strength U. (b) Density of states spectra for a clean system.

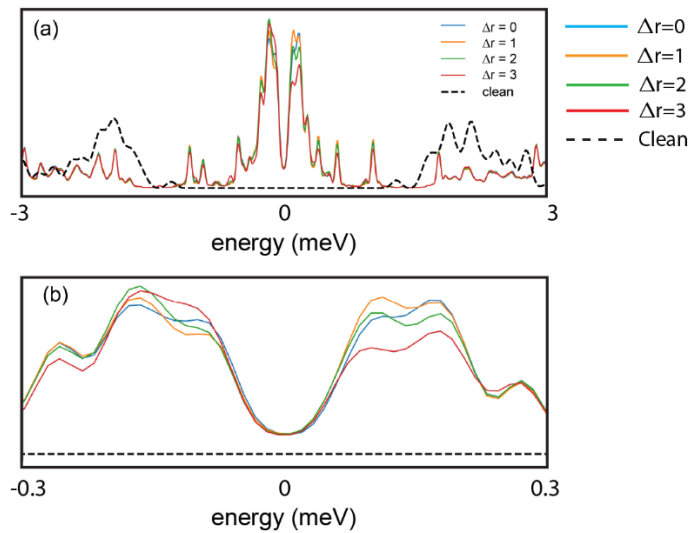

**Figure S11. Theoretical density of states near a π-phase domain wall**. (a) Orbitally-summed N(**r**,E) at a distance Δr from the domain wall. The local density of states for a clean system (dashed line) is provided as a guide to the eye. (b) The result in (a), magnified near zero energy.

## Supplementary Section 6

To further verify the role of non-trivial topology in creating the domain wall modes, we have studied twin-domain walls in the topologically trivial but related superconducting compound, FeSe. We find that twin domain walls in FeSe suppress the superconducting order parameter but do not result in a constant DOS.

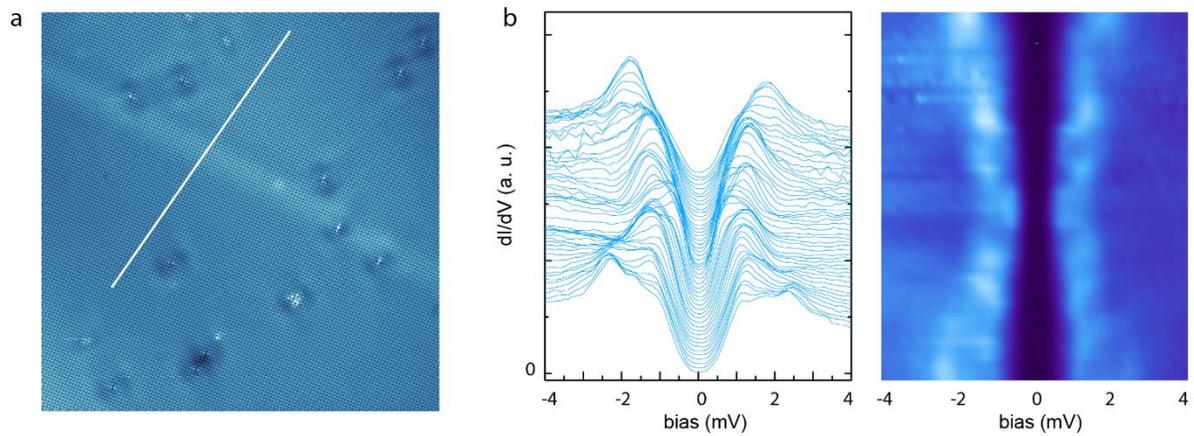

**Figure S12. Twin boundary in FeSe.** We find that twin domain walls in FeSe suppress the superconducting order parameter but do not result in a constant DOS